\documentclass[twocolumn,english,aps,prd,nofootinbib]{revtex4-2}
\usepackage{lmodern}
\usepackage[LGR,T1]{fontenc}
\usepackage[latin9]{inputenc}
\usepackage{xcolor}
\usepackage{babel}
\usepackage{amsmath}
\usepackage{amssymb}
\usepackage{graphicx}
\usepackage[unicode=true]
 {hyperref}

\makeatletter

\DeclareRobustCommand{\greektext}{%
  \fontencoding{LGR}\selectfont\def\encodingdefault{LGR}}
\DeclareRobustCommand{\textgreek}[1]{\leavevmode{\greektext #1}}
\ProvideTextCommand{\~}{LGR}[1]{\char126#1}

\usepackage{enumerate}

\makeatother

\begin{document}
\title{Violation of $\gamma$ in Brans-Dicke gravity}
\author{Hoang Ky Nguyen$\,$}
\email[\ ]{hoang.nguyen@ubbcluj.ro}

\affiliation{Department of Physics, Babe\c{s}-Bolyai University, Cluj-Napoca 400084,
Romania}
\author{Bertrand Chauvineau$\,$}
\email[\ ]{bertrand.chauvineau@oca.eu}

\affiliation{Universit\'e C\^ote d'Azur, Observatoire de la C\^ote d\textquoteright Azur,
CNRS, Laboratoire Lagrange, Nice cedex 4, France}
\date{June 25, 2024}
\begin{abstract}
\vskip2pt The Brans Class I solution in Brans-Dicke gravity is a
staple in the study of gravitational theories beyond General Relativity.
Discovered in 1961, it describes the exterior vacuum of a spherical
Brans-Dicke star and is characterized by two adjustable parameters.
Surprisingly, the relationship between these parameters and the properties
of the star has not been rigorously established. In this Proceeding,
we bridge this gap by deriving \emph{the} complete exterior solution
of Brans Class I, expressed in terms of the total energy and total
pressure of the spherisymmetric gravity source. The solution allows
for the \emph{exact} derivation of \emph{all} post-Newtonian parameters
in Brans-Dicke gravity for far field regions of a spherical source.
Particularly for the $\gamma$ parameter, instead of the conventional
result $\gamma_{\,\text{PPN}}=\frac{\omega+1}{\omega+2}$, we obtain
the analytical expression $\gamma_{\,\text{exact}}=\frac{\omega+1+(\omega+2)\,\Theta}{\omega+2+(\omega+1)\,\Theta}$
where $\Theta$ is the ratio of the total pressure $P_{\parallel}^{*}+2P_{\perp}^{*}$
and total energy $E^{*}$ contained within the mass source. Our \emph{non-perturbative}
$\gamma$ formula is valid for all field strengths and types of matter
comprising the mass source. Consequently, observational constraints
on $\gamma$ thus set \emph{joint} bounds on $\omega$ and $\varTheta$,
with the latter representing a global characteristic of the mass source.
More broadly, our formula highlights the importance of pressure (when
$\varTheta\neq0$) in spherical Brans-Dicke stars, and potentially
in stars within other modified theories of gravitation.
\end{abstract}
\maketitle
\emph{Background}---Brans--Dicke gravity is the second most studied
theory of gravitation besides General Relativity. It represents one
of the simplest extensions of gravitational theory beyond GR \citep{BransDicke-1961}.
It is characterized by an additional dynamical scalar field $\phi$
which, in the original vision of Brans and Dicke in 1961, acts like
the inverse of a variable Newton `constant' $G$. The scalar field
has a kinetic term, governed by a (Brans-Dicke) parameter $\omega$
in the following gravitation action
\begin{equation}
S=\frac{1}{16\pi}\int d^{4}x\sqrt{-g}\left[\Phi\,\mathcal{R}-\frac{\omega}{\Phi}g^{\mu\nu}\partial_{\mu}\Phi\partial_{\nu}\Phi\right]
\end{equation}
In the limit of infinite value for $\omega$, the kinetic term is
generally said to be `frozen', rendering $\Phi$ being a constant
value everywhere. In this limit, if the field $\phi$ approaches its
(non-zero) constant value in the rate $\mathcal{O}\left(1/\omega\right)$,
the term $\frac{\omega}{\Phi}g^{\mu\nu}\partial_{\mu}\Phi\partial_{\nu}\Phi$
would approach zero at the rate $\mathcal{O}\left(1/\omega\right)$
and hence become negligible compared with the term $\Phi\,\mathcal{R}$,
effectively recovering the classic Einstein--Hilbert action. \footnote{It has been shown that for non-static and/or in the presence of singularity,
the rate of convergence is $\mathcal{O}\left(1/\sqrt{\omega}\right)$.
This topic is beyond the scope of this Proceeding however, as we shall
only consider a static and regular case here. For more information,
we refer the reader to our recent work \citep{Nguyen-2023-BDKG},
where we also reviewed the literature on the $\mathcal{O}\left(1/\sqrt{\omega}\right)$
anomaly.}\vskip4pt

Together with its introduction \citep{BransDicke-1961}, Brans also
identified four classes of exact solutions in the static spherically
symmetric (SSS) setup \citep{Brans-1962}. The derivation of the Brans
solutions was explicitly carried out by Bronnikov in 1973 \citep{Bronnikov-1973}.
Of the four classes, only the Brans Class I is physically meaningful,
however. It can recover the Schwarzschild solution in its parameter
space.\vskip4pt

For comparison with observations or experiments, Brans derived the
Robertson (or Eddington-Robertson- Schiff) $\beta$ and $\gamma$
post-Newtonian (PN) parameters based on his Class I solution: 
\begin{align}
\beta_{\,\text{PPN}} & =1\label{eq:PPN-beta}\\
\gamma_{\,\text{PPN}} & =\frac{\omega+1}{\omega+2}\label{eq:PPN-gamma}
\end{align}
The $\gamma$ parameter is important as it governs the amount of space-curvature
produced by a body at rest and can be directly measured via the detection
of light deflection and the Shapiro time delay. The parametrized post-Newtonian
(PPN) $\gamma$ formula recovers the result $\gamma_{\,\text{GR}}=1$
known for GR in the limit of infinite $\omega$, in which the BD scalar
field becomes constant everywhere. Current bounds using Solar System
observations set the magnitude of $\omega$ to exceed $40,000$ \citep{Will2}.\vskip4pt

We should emphasize that the ``conventional'' results \eqref{eq:PPN-beta}
and \eqref{eq:PPN-gamma} were derived under \emph{the assumption
of zero pressure in the gravity source}. It should be noted that these
formulae can also be deduced directly from the PPN formalism for the
Brans--Dicke action, without resorting to the Brans Class I solution
\citep{Weinberg,Will1}. The PPN derivation relies on two crucial
approximations: (i) weak field and (ii) slow motions. Regarding the
latter approximation, an often under-emphasized point is that not
only must the stars be in slow motion, but the microscopic constituents
that comprise the stars must also be in slow motion. This translates
to the requirement that the matter inside the stars exert low pressure,
characterizing them as ``Newtonian'' stars. \vskip4pt

The purpose of our paper is two-fold. Firstly, the analytical form
for the exterior vacuum contains two adjustable parameters. The issue
in determining them from the energy and pressure profiles inside the
mass source has not been rigorously addressed in the literature. Establishing
their relationships with the mass source would typically require the
full machinery of the Tolman--Oppenheimer--Volkoff (TOV) equations
tailored for Brans--Dicke gravity \citep{Nguyen-compact-star-2}.
Moreover, solving the TOV equations, even in the simpler theory of
GR, generally requires numerical methods except for a few isolated,
unrealistic cases such as incompressible fluids. Therefore, at first
glance, deriving a concrete expression for these relationships might
seem elusive. Surprisingly, as we shall show in this Proceeding, this
view is overly pessimistic. It turns out that the full machinery of
the TOV equation is not necessary. Instead, only a subset of the field
equation and the scalar equation of BD will be needed. This is because
only two equations are required to fix the two free parameters of
the exterior vacuum. We shall present a rigorous yet parsimonious
derivation, which only became available through our recent publication
\citep{2024-gamma-PLB}.\vskip4pt

Secondly, the complete solution enables the derivation of \emph{any}
PN parameters applicable for far-field regions in static spherical
Brans-Dicke stars. As we shall show in this Proceeding, the derivation
is \emph{non-perturbative} and \emph{avoids the two PPN approximations}
requiring the weak field and the low pressure mentioned above. \vskip4pt

The material presented in this Proceeding was developed during the
preparation of our two recent papers \citep{2024-gamma-PLB,2024-gamma-EPJC}.
For a more detailed exposition of the conceptualization and technical
points, we refer the reader to these papers.\vskip12pt

\emph{The field equations and the energy-momentum tensor}---It is
well documented \citep{Bronnikov-1973} that upon the Weyl mapping
$\bigl\{\tilde{g}_{\mu\nu}:=\Phi\,g_{\mu\nu}$, $\tilde{\Phi}:=\ln\Phi\bigr\}$,
the gravitational sector of the BD action can be brought to the Einstein
frame as $\int d^{4}x\frac{\sqrt{-\tilde{g}}}{16\pi}\Bigl[\tilde{\mathcal{R}}-\left(\omega+3/2\right)\tilde{\nabla}^{\mu}\tilde{\Phi}\tilde{\nabla}_{\mu}\tilde{\Phi}\Bigr]$.
The Einstein-frame BD scalar field $\tilde{\Phi}$ has a kinetic term
with a signum determined by $(\omega+3/2)$. Unless stated otherwise,
we shall restrict our consideration to the normal (``non-phantom'')
case of $\omega>-3/2$, where the kinetic energy for $\tilde{\Phi}$
is positive.\vskip8pt

The field equations are
\begin{align}
R_{\mu\nu}-\frac{\omega}{\Phi^{2}}\partial_{\mu}\Phi\partial_{\nu}\Phi-\frac{1}{\Phi}\partial_{\mu}\partial_{\nu}\Phi+\Gamma_{\mu\nu}^{\lambda}\partial_{\lambda}\ln\Phi\ \ \ \ \ \ \ \ \ \ \nonumber \\
=\frac{8\pi}{\Phi}\left(T_{\mu\nu}-\frac{\omega+1}{2\omega+3}T\,g_{\mu\nu}\right)\label{eq:R-eqn}\\
\partial_{\mu}\left(\sqrt{-g}\,g^{\mu\nu}\partial_{\nu}\Phi\right)=\frac{8\pi}{2\omega+3}T\sqrt{-g}\ \ \ \ \ \ \ \ \,\ \ \ \ \ \ \ \label{eq:Phi-eqn}
\end{align}
In the isotropic coordinate system which is static and spherically
symmetric, the metric can be written as
\begin{equation}
ds^{2}=-A(r)dt^{2}+B(r)\left[dr^{2}+r^{2}\left(d\theta^{2}+\sin^{2}\theta d\varphi^{2}\right)\right]\label{eq:spherical metric}
\end{equation}
It is straightforward to verify, from Eqs. \eqref{eq:R-eqn}--\eqref{eq:spherical metric},
that the most general form for the energy-momentum tensor (EMT) in
this setup is
\begin{align}
T_{\mu}^{\nu} & =\text{diag}\bigl(-\epsilon,\,p_{\Vert},\,p_{\bot},\,p_{\bot}\bigr)\label{eq:stress tensor}
\end{align}
where the energy density $\epsilon$, the radial pressure $p_{\parallel}$
and the tangential pressure $p_{\perp}$ are functions of $r$. Note
that the EMT is anisotropic if $p_{\parallel}\neq p_{\perp}$. The
trace of the EMT is
\begin{equation}
T=-\epsilon+p_{\parallel}+2p_{\perp}\,.\label{eq:EMT-trace}
\end{equation}
\vskip12pt

\emph{The Brans Class I vacuum solution outside a star}---It is known
that the scalar--metric for the vacuum is the Brans Class I solution
(which satisfies Eqs. \eqref{eq:Phi-eqn}--\eqref{eq:R-eqn} for
$T_{\mu\nu}=0$) \citep{Brans-1962}. In the isotropic coordinate
system \eqref{eq:spherical metric}, the solution reads \citep{Brans-1962}
\begin{equation}
\left\{ \begin{array}{l}
A=\left(\frac{r-k}{r+k}\right)^{\frac{2}{\lambda}}\\
B=\left(1+\frac{k}{r}\right)^{4}\left(\frac{r-k}{r+k}\right)^{2-2\frac{\Lambda+1}{\lambda}}\\
\Phi=\left(\frac{r-k}{r+k}\right)^{\frac{\Lambda}{\lambda}}
\end{array}\right.\ \ \ \ \ \text{for }r\geqslant r_{*}\label{eq:BC1}
\end{equation}
where $r_{\ast}$ is the star's radius, and 
\begin{equation}
\lambda^{2}=\left(\Lambda+1\right)^{2}-\Lambda\left(1-\frac{\Lambda}{2}\omega\right)\label{eq:relation}
\end{equation}
Since $\lambda$ and $\Lambda$ are linked by \eqref{eq:relation},
this solution involves two independent parameters, which one chooses
to be $\left(k,\,\Lambda\right)$.\vskip12pt

\emph{The field equations in the interior---}For the region $r\leqslant r_{*}$,
substituting metric \eqref{eq:spherical metric} and the BD field
$\Phi(r)$ into Eq. \eqref{eq:Phi-eqn} and the \emph{$00-$component}
of Eq. \eqref{eq:R-eqn} and using the EMT in Eq. \eqref{eq:stress tensor},
the functions $A(r)$, $B(r)$, $\Phi(r)$ satisfy the 2 following
ordinary differential equations (ODEs): 
\begin{align}
\left(r^{2}\sqrt{AB}\Phi^{\prime}\right)^{\prime} & =\frac{8\pi}{2\omega+3}\Bigl[-\epsilon+p_{\Vert}+2p_{\bot}\Bigr]r^{2}\sqrt{AB^{3}}\label{eq:BD eq(Phi)}\\
\left(r^{2}\Phi\sqrt{\frac{B}{A}}A^{\prime}\right)^{\prime} & =16\pi\Bigl[\,\epsilon+\frac{\omega+1}{2\omega+3}\bigl(-\epsilon+p_{\Vert}+2p_{\bot}\bigr)\Bigr]r^{2}\sqrt{AB^{3}}\label{eq:BD eq(A)}
\end{align}

These equations offer the advantage of having both their left hand
sides in exact derivative forms. Let us integrate Eqs. \eqref{eq:BD eq(Phi)}
and \eqref{eq:BD eq(A)} from the star's center, viz. $r=0$, to a
coordinate $r>r_{\ast}$. The $\left(A,B,\Phi\right)$ functions are
then given by \eqref{eq:BC1} at $r$. For $r>r_{*}$, both $r^{2}\sqrt{AB}\Phi^{\prime}$
and $r^{2}\Phi\sqrt{\frac{B}{A}}A^{\prime}$ terms that enter the
left hand sides of \eqref{eq:BD eq(Phi)} and \eqref{eq:BD eq(A)}
are $r-$independent, since the right hand sides of these equations
vanish in the \emph{exterior} vacuum. On the other hand, regularity
conditions inside the star impose $\Phi^{\prime}\left(0\right)=A^{\prime}\left(0\right)=0$
(i.e. no conic singularity) and finite values of the fields themselves.
The calculation yields
\begin{align}
\frac{k\,\Lambda}{\lambda} & =\frac{4\pi}{2\omega+3}\int_{0}^{r_{\ast}}dr\,r^{2}\sqrt{AB^{3}}\Bigl[-\epsilon+p_{\Vert}+2p_{\bot}\Bigr]\label{eq:BD eq(Phi) integr}
\end{align}
and
\begin{align}
\frac{k}{\lambda} & =\frac{4\pi}{2\omega+3}\int_{0}^{r_{\ast}}dr\,r^{2}\sqrt{AB^{3}}\times\nonumber \\
 & \ \ \ \ \ \ \ \ \ \ \ \ \ \ \ \ \ \Bigl[(\omega+2)\epsilon+(\omega+1)\bigl(p_{\Vert}+2p_{\bot}\bigr)\Bigr]\,.\label{eq:BD eq(A) integr}
\end{align}
Let us note that $r^{2}\sqrt{AB^{3}}$ is the square root of the determinant
of the metric, up to the $\sin\theta$ term. (Accordingly, the integrals
in the right hand sides of Eqs. \eqref{eq:BD eq(Phi) integr} and
\eqref{eq:BD eq(A) integr} are invariant through radial coordinate
transformations, since the combination $r^{2}\sqrt{AB^{3}}\sin\theta$
is equivalent to $\sqrt{-g}$.) We then can define the energy's and
pressures' integrals by\vspace{-0.25cm}
\begin{eqnarray}
E^{\ast} & = & 4\pi\int_{0}^{r_{\ast}}dr\,r^{2}\sqrt{AB^{3}}\,\epsilon\label{eq:star's energy integr}\\
P_{\Vert}^{\ast} & = & 4\pi\int_{0}^{r_{\ast}}dr\,r^{2}\sqrt{AB^{3}}\,p_{\parallel}\label{eq:star's rad pressure integr}\\
P_{\bot}^{\ast} & = & 4\pi\int_{0}^{r_{\ast}}dr\,r^{2}\sqrt{AB^{3}}\,p_{\perp}\,.\label{eq:star's orthog pressure integr}
\end{eqnarray}
Inserting in \eqref{eq:BD eq(Phi) integr} and \eqref{eq:BD eq(A) integr},
we obtain
\begin{equation}
\frac{k}{\lambda}=E^{*}\Bigl[\frac{\omega+2}{2\omega+3}+\frac{\omega+1}{2\omega+3}\,\Theta\Bigr]\label{eq:k-formula}
\end{equation}
and
\begin{equation}
\Lambda=\frac{\Theta-1}{\omega+2+(\omega+1)\,\Theta}\label{eq:Lambda-formula}
\end{equation}
in which the dimensionless parameter $\Theta$ is defined as
\begin{equation}
\Theta:=\frac{P_{\parallel}^{*}+2\,P_{\perp}^{*}}{E^{*}}\,.\label{eq:Theta-def}
\end{equation}
Together with \eqref{eq:BC1} and \eqref{eq:relation}, these expressions
provide a complete expression for the exterior spacetime and scalar
field of a spherical BD star. To the best of our knowledge, this prescription
was not made explicitly documented in the literature, until our recent
works \citep{2024-gamma-EPJC,2024-gamma-PLB}.\vskip4pt

For a perfect fluid, $p_{\Vert}=p_{\bot}\equiv p$, thence $P_{\Vert}^{\ast}=P_{\bot}^{\ast}\equiv P$.
The equations \eqref{eq:relation}, \eqref{eq:BD eq(Phi) integr}
and \eqref{eq:BD eq(A) integr} fully determine the exterior solution
\eqref{eq:BC1} once the integrals \eqref{eq:star's energy integr}--\eqref{eq:star's orthog pressure integr}
are known, with these integrals being fixed by the stellar internal
structure model. This explicitly determines the particles' motion
outside the star, in both the remote and close to the star regions.

$ $\vskip4pt

\emph{The ($\beta$, $\gamma$, $\delta$) PN parameters}---In remote
spatial regions, a static spherically symmetric metric in isotropic
coordinates can be expanded as \citep{Weinberg}:
\begin{align}
ds^{2} & =-\left(1-2\,\frac{M}{r}+2\beta\,\frac{M^{2}}{r^{2}}+\dots\right)dt^{2}\nonumber \\
 & +\left(1+2\gamma\,\frac{M}{r}+\frac{3}{2}\delta\,\frac{M^{2}}{r^{2}}+\dots\right)\left(dr^{2}+r^{2}d\Omega^{2}\right)\label{eq:taylor}
\end{align}
in which $\beta$ and $\gamma$ are the Robertson (or Eddington-Robertson-Schiff)
parameters, whereas $\delta$ is the second-order PN parameter (for
both light and planetary like motions). It is straightforward to verify
that the Schwarzschild metric yields 
\begin{equation}
\beta_{\,\text{Schwd}}=\gamma_{\,\text{Schwd}}=\delta_{\,\text{Schwd}}=1
\end{equation}
The metric in Eq. \eqref{eq:BC1} can be re-expressed in the expansion
form\begin{widetext}
\begin{align}
ds^{2} & =-\left(1-\frac{4}{\lambda}\frac{k}{\rho}+\frac{8}{\lambda^{2}}\frac{k^{2}}{r^{2}}+\dots\right)dt^{2}+\left(1+\frac{4}{\lambda}(1+\Lambda)\frac{k}{r}+\frac{2}{\lambda^{2}}\left(4(1+\Lambda)^{2}-\lambda^{2}\right)\frac{k^{2}}{r^{2}}+\dots\right)\left(dr^{2}+r^{2}d\Omega^{2}\right)\label{eq:Robertson}
\end{align}
\end{widetext}Comparing Eq. \eqref{eq:taylor} against Eq. \eqref{eq:Robertson}
and setting
\begin{equation}
M=2\,\frac{k}{\lambda}\label{eq:M-def}
\end{equation}
we obtain
\begin{align}
\beta_{\,\text{exact}} & \ =\ 1\label{eq:beta-exact-0}\\
\gamma_{\,\text{exact}} & \ =\ 1+\Lambda\label{eq:gamma-exact-0}\\
\delta_{\,\text{exact}} & \ =\ \frac{1}{3}\left(4(1+\Lambda)^{2}-\lambda^{2}\right)\label{eq:delta-exact-0}
\end{align}
where we have used the subscript ``exact'' as emphasis. Note that
$\Lambda$ directly measures the deviation of the $\gamma$ parameters
from GR ($\gamma_{\text{GR}}=1$). From Eq. \eqref{eq:Lambda-formula},
\emph{$\Lambda$ depends on both $\omega$ and $\Theta$}. Finally,
we arrive at
\begin{align}
\gamma_{\,\text{exact}}\  & =\ \frac{\omega+1+(\omega+2)\hspace{0.02cm}\varTheta}{\omega+2+(\omega+1)\hspace{0.02cm}\varTheta}\label{eq:gamma-exact-1}
\end{align}
which can also be conveniently recast as
\begin{align}
\gamma_{\,\text{exact}}\  & =\ \frac{\gamma_{\,\text{PPN}}+\Theta}{1+\gamma_{\,\text{PPN}}\,\Theta}\label{eq:gamma-exact-2}
\end{align}
by recalling that $\gamma_{\,\text{PPN}}=\frac{\omega+1}{\omega+2}$.
To our knowledge, the closed-form expression \eqref{eq:gamma-exact-1}
for $\gamma$ was absent in the literature, until our recent works
\citep{2024-gamma-EPJC,2024-gamma-PLB}.\vskip8pt

\begin{figure}[!t]
\noindent \begin{centering}
\includegraphics[scale=0.6]{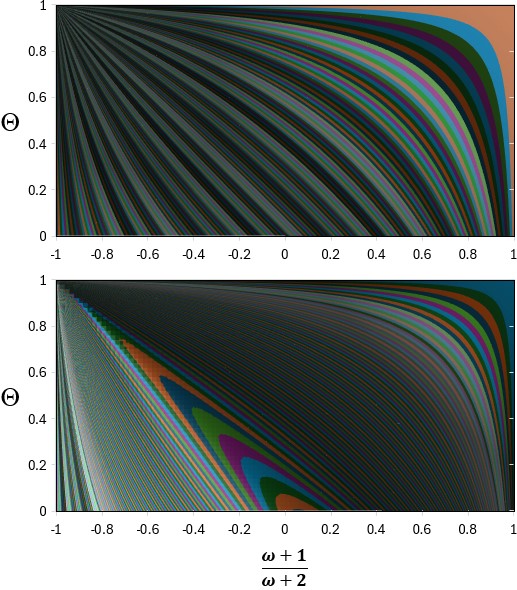}
\par\end{centering}
\caption{\label{fig:gamma-countour}Contour plots of $\gamma_{\,\text{exact}}$
(upper panel) and $\delta_{\,\text{exact}}$ (lower panel) in terms
of $\varTheta$ and $\frac{\omega+1}{\omega+2}$, for the range of
$\varTheta\in[0,1]$ and $\omega\in(-3/2,+\infty)$, the latter corresponding
to $\frac{\omega+1}{\omega+2}\in(-1,1)$. A measured $\gamma_{\,\text{exact}}\approx1$
could mean $\frac{\omega+1}{\omega+2}\approx1$ (i.e., $\omega\gg1$)
or $\varTheta\approx1$ (i.e., ultra-relativistic matter). Contours
are equally spaced in 0.01 increment. For a given contour in the upper
panel, the corresponding value of $\gamma_{\,\text{exact}}$ can be
read on the abscissa axis, where the contour intersects it. By measuring
\emph{both} $\gamma$ and $\delta$, the values of $\omega$ and $\Theta$
may be determined.}
\end{figure}

Regarding $\delta$:
\begin{align}
\delta_{\,\text{exact}} & =\frac{1}{\left[\omega+2+(\omega+1)\,\Theta\right]^{2}}\Bigl[\Bigl(\omega^{2}+\frac{3}{2}\omega+\frac{1}{3}\Bigr)\nonumber \\
 & +\Bigl(2\omega^{2}+\frac{19}{3}\omega+\frac{13}{3}\Bigr)\Theta+\Bigl(\omega^{2}+\frac{25}{6}\omega+\frac{13}{3}\Bigr)\Theta^{2}\Bigr]\label{eq:delta-exact-1}
\end{align}
Figure \ref{fig:gamma-countour} shows contour plots of $\gamma_{\,\text{exact}}$
and $\delta_{\,\text{exact}}$ as functions of $\gamma_{\,\text{PPN}}$
(i.e, $\frac{\omega+1}{\omega+2}$) and $\Theta$. In addition, with
the aid of Eqs. \eqref{eq:k-formula} and \eqref{eq:Theta-def}, Eq.
\eqref{eq:M-def} produces the active gravitational mass
\begin{align}
M & =\frac{2\omega+4}{2\omega+3}\,E^{*}+\frac{2\omega+2}{2\omega+3}\,\left(P_{\parallel}^{*}+2P_{\bot}^{*}\right)\label{eq:active-mass}
\end{align}
where the contribution of pressure to the active gravitational mass
is evident \citep{Baez-2005,Ehlers-2005}.\vskip12pt

\emph{Degeneracy at ultra-high pressure}---For $\Theta\rightarrow1^{-}$,
both $\gamma$ and $\delta$ go to 1, their GR counterpart values.
Generally speaking, for $\Theta\rightarrow1^{-}$, since $\Lambda\rightarrow0$
and $\lambda\rightarrow1$ regardless of $\omega$ (provided that
$\omega\in(-3/2,+\infty)$), the value of $k$ approaches 
\begin{equation}
k\rightarrow\frac{\omega+2}{2\omega+3}\,E^{*}+\frac{\omega+1}{2\omega+3}\,\left(P_{\parallel}^{*}+2P_{\bot}^{*}\right)
\end{equation}
The $\omega-$dependence is thus absorbed into $k$, and the Brans
Class I solution degenerates to the Schwarzschild solution

\begin{equation}
\left\{ \begin{array}{l}
A=\left(\frac{r-k}{r+k}\right)^{2}\\
B=\left(1+\frac{k}{r}\right)^{4}\\
\Phi=1
\end{array}\right.\ \ \ \ \ \text{for }r\geqslant r_{*}
\end{equation}
Therefore, ultra-relativistic Brans-Dicke stars are \emph{indistinguishable}
from their GR counterparts, as far as their exterior vacua are concerned.
This fact can be explained by the following observation: For ultra-relativistic
matter, the trace of the EMT vanishes, per Eq. \eqref{eq:EMT-trace}.
The scalar equation \eqref{eq:Phi-eqn} then simplifies to $\square\,\Phi=0$
\emph{everywhere}. Coupled with the regularity condition at the star
center, this ensures a constant $\Phi$ throughout the spacetime which
is now described by the Schwarzschild solution. Consequently, the
scalar degree of freedom in BD gravity is suppressed in the ultra-relativistic
limit. This prompts an intriguing possibility whether Birkhoff's theorem
is fully restored in this limit.\vskip12pt

\emph{Discussions}---Formulae \eqref{eq:gamma-exact-1} and \eqref{eq:delta-exact-1}
are the essential outcome of this Proceeding:\vskip4pt
\begin{itemize}
\item \emph{Non-perturbative approach}: $\ $Our derivation is non-perturbative
in nature. It makes use of the \emph{integrability} of the $00-$component
of the field equation \eqref{eq:R-eqn}, along with the scalar field
equation \eqref{eq:Phi-eqn}.
\item \emph{Parsimony}: $\ $Our derivation relies solely on the scalar
field equation and the 00-component of the field equation, without
the need for the full set of equations, specifically the $11-$ and
$22-$ components of the field equation \footnote{Note that establishing the functional form of the Brans Class I solution
still requires the full set of equations.}. The additional physical assumptions employed are the regularity
at the star's center and the existence of the star's surface separating
the interior and the exterior. 
\item \emph{Universality of results}: $\ $The final formulae, \eqref{eq:gamma-exact-1}
and \eqref{eq:delta-exact-1}, hold for all field strengths and all
types of matter (whether convective or non-convective, for example).
We do not assume the matter comprising the stars to be a perfect fluid
or isentropic.
\item \emph{Higher-derivative characteristics}: $\ $In contrast to the
one-parameter Schwarzschild metric, the Brans Class I solution depends
on two parameters, i.e. the solution is not only defined by its gravitational
mass, but also by \emph{a scalar mass} besides the gravitational one
\citep{Bronnikov-1973}. The exterior BD vacuum should reflect the
internal structure and composition of the star. This expectation is
confirmed in Eqs. \eqref{eq:gamma-exact-1} and \eqref{eq:delta-exact-1},
highlighting the role of the parameter $\Theta$.
\item \emph{Role of pressure}: $\ $Figure \ref{fig:gamma-countour} shows
contour plots of $\gamma_{\,\text{exact}}$ and $\delta_{\,\text{exact}}$
as functions of $\frac{\omega+1}{\omega+2}$ and $\Theta$. There
are three interesting observations:
\begin{itemize}
\item An ultra-relativistic limit, $\varTheta\simeq1^{-}$, would render
$\gamma_{\,\text{exact}}\simeq1$, \emph{regardless} of $\omega$.
\item For Newtonian stars, i.e. low pressure ($\varTheta\approx0$), the
PPN result is a good approximation \emph{regardless} of the field
strength.
\item A joint measurement of $\gamma$ and $\delta$ in principle can determine
$\omega$ and $\Theta$. However, due to the non-linear relationships
in \eqref{eq:gamma-exact-1} and \eqref{eq:delta-exact-1}, for a
given pair of $\{\gamma,\ \delta\}$, multiple solutions for $\{\omega,\ \Theta\}$
can exist. A measurement of a third PN parameter (apart from $\beta$)
in principle can resolve the multiplicity problem.
\end{itemize}
\end{itemize}
\emph{\indent Conclusion}---We have derived the exact analytical
formulae, \eqref{eq:gamma-exact-1} and \eqref{eq:delta-exact-1},
for the PN parameters $\gamma$ and $\delta$ for spherical mass sources
in BD gravity. The derivation relies on the integrability of the $00-$component
of the field equation, rendering it non-perturbative and applicable
for any field strength and type of matter constituting the source.
The conventional PPN result for BD gravity $\gamma_{\,\text{PPN}}=\frac{\omega+1}{\omega+2}$
lacks dependence on the physical features of the mass source. In the
light of our exact results, the $\gamma_{\,\text{PPN}}$ should be
regarded as an approximation for stars in modified gravity under low-pressure
conditions. Our findings expose the limitations of the PPN formalism,
particularly in scenarios characterized by high star pressure. It
is reasonable to expect that the role of pressure may extend to other
modified theories of gravitation.\vskip12pt

\emph{Acknowledgments}---BC thanks Antoine Strugarek for helpful
correspondences. HKN thanks Mustapha Azreg-A\"inou, Valerio Faraoni,
Tiberiu Harko, Viktor Toth, and the participants of the XII Bolyai--Gauss--Lobachevsky
Conference (BGL-2024): Non-Euclidean Geometry in Modern Physics and
Mathematics (Budapest, May 1-3, 2024) for valuable commentaries.

\end{document}